\documentclass[aps,twocolumn,showpacs,preprintnumbers,amsmath,amssymb,showkeys,floatfix,superscriptaddress]{revtex4}

\usepackage{graphicx,epsf,epsfig}
\usepackage{dcolumn}
\usepackage{bm}
\usepackage{color}
\usepackage{ulem}

\begin{document}


\title{Reversal of current blockade through multiple trap correlations.}
\author{Jack Chan}
\author{Brian Burke}
\author{Kenneth Evans}
\author{Keith A. Williams}
\affiliation{Department of Physics, University of Virginia, Charlottesville, VA 22904}
\author{Smitha Vasudevan}
\author{Mingguo Liu}
\author{Joe Campbell}
\author{Avik W. Ghosh}
\affiliation{Department of Electrical and Computer Engineering\\
University of Virginia, Charlottesville, VA 22904}
\email[Corresponding Author: ]{kwilliams@virginia.edu}

\begin{abstract}
Current noise in electronic devices usually arises from uncorrelated charging events, with individual transitions resolved only at low temperatures. However, in 1-D nanotube-based transistors, we have observed random telegraph signal (RTS) with unprecedented signal-to-noise ratio at room temperature.  In addition, we find evidence for cooperative multi-trap interactions that give rise to a characteristically terminated RTS: current blockade induced by one trap is found to {\it{fully reverse}} through electrostatic `passivation' by another. Our observations are well described by a robust quantum transport model that demonstrates how strong correlation and fast varying potentials can resolve energetically proximal states along a 1-D channel.
\end{abstract}

\pacs{Valid PACS appear here}
\maketitle


\textcolor{black}{The transport channels of micro-electronic devices are trending to ever larger surface-to-volume ratios, making them increasingly susceptible to stochastic effects such as random telegraph signals (RTS) and 1/f noise arising from dopant fluctuations, dangling bonds and charge traps at surfaces and interfaces.} RTS effects in bulk Si-MOSFETs have been studied for several decades \cite{kirton}, \textcolor{black}{albeit} at temperatures of a few Kelvin. \textcolor{black}{Lower dimensional quantum confinement amplifies the effects of individual defects through stronger wave-function overlap}. \textcolor{black}{For instance,} \textcolor{black}{individual defects in} carbon nanotubes lead to prominent RTS, well resolved at \textcolor{black}{larger (though not room)} temperatures \cite{zhou1},\cite{perkins}, because the charge traps can completely block current flow. \textcolor{black}{More significantly, the amplified electronic interaction
in lower-dimension can lead to novel correlation physics detectable at room temperature}.

\begin{figure}[h!]
\hskip 0.2cm\centerline{\epsfxsize=3.4in\epsfbox{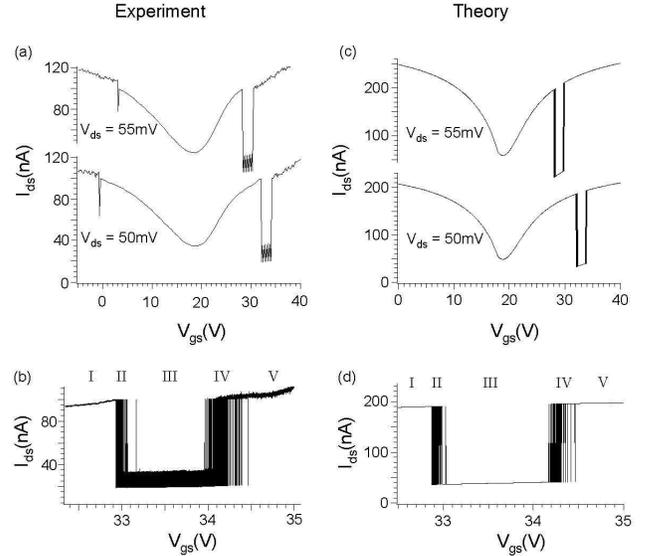}}
\vskip -0.4cm
\caption{{\it{(a) $I_{ds}-V_{gs}$ plots for $V_{ds}= 50$ mV and 55 mV, measured at room temperature under vacuum. Five different transport regimes are identified with numerals. Scan rate and scan resolution are 80 V/sec and 6.25 data points/V respectively. (b) Higher resolution scan of $I_{ds}$ versus $V_{gs}$ at the blockade window for $V_{ds}= 50$ mV. Scan rate and scan resolution are 0.125 V/sec and 4000 data point/V respectively. (c,d) NEGF simulation of current flow through CNT-FET coupled with Monte Carlo model for trap-assisted scattering.}}}
\label{ivs}
\end{figure}

\textcolor{black}{In this letter, we report RTS signatures in \textcolor{black}{long} \textcolor{black}{carbon nanotube field-effect transistors} (CNT-FETs) with channel length of \textcolor{black}{$20 \mu m$}, that are quite unique in many aspects, (a)} \textcolor{black}{exhibiting well-resolved} amplitudes of up to $ 80 \%$ of the baseline ambipolar current \textcolor{black}{even at room temperature} (Fig. 1a). \textcolor{black}{(b)} \textcolor{black}{Further,} we demonstrate how \textcolor{black} {correlated} scattering via adjacent traps along the CNT introduces a characteristically truncated, gateable Coulomb blockade window flanked by RTS on either side, amounting to the near-complete blocking and subsequent unblocking of CNT conduction (Fig. 1b). \textcolor{black}{(c) Our experimental observations are explained using a quantum transport model (Fig. 1c, d) for current flow coupled with a self-consistent Coulomb potential for multi-trap interaction and a Monte Carlo simulation of stochastic switching. Our model clearly illustrates how \textcolor{black}{a logarithmically varying capacitance} (Fig. 3) allows a distant gate \textcolor{black}{to} tune the blockade and resolve the cooperative effects of two nearby traps that annihilate each other (Fig. 2)}.

The fabrication process of a typical CNT-FET inevitably introduces defects on the CNT, in the oxide, or at their interfaces \cite{zhou3}, through the high-temperature annealing of $SiO_2$ in $H_2$ \cite{zhang} and the interaction with high energy electrons from electron beam microscopy \cite{pelz}.
These defects can trap charges which may subsequently interact electrostatically with the CNT carriers. The
stochastic trapping and detrapping of carriers near resonance with the Fermi level $E_F$ creates a
commensurate flicker in the output current in the form of RTS \cite{tersoff,zhou4,golovchenko,zhou5}.


{\it{Experiment.}} CNT-FETs were fabricated on B-doped Si(100) substrates with a 500 nm thick thermal oxide and resistivity of $0.001-0.0025 \Omega cm^{-1}$ (SQI). Catalysts were deposited on pre-patterned sites and CNTs were grown on the oxide surface by chemical vapor deposition (CVD) \cite{dai}. After annealing with $H_2$ at $ 900^{\circ}$C for 10 minutes, CVD synthesis is carried out for 15 minutes using methane and hydrogen at the same temperature. Ti/Au electrodes were deposited by photolithography over selected, isolated CNTs to introduce source and drain contacts after analysis by SEM. The contacts were then annealed in Argon $99.999 \%$, GT\&S) at atmospheric pressure and at a temperature of $~600^{\circ}$C for 5 minutes \cite{iijima3}, \cite{martel}. As usual for these prototype devices, the conductance through the CNT channels is modulated by a universal backgate (doped poly-Si). With the device contained in a variable temperature probe station (TTP-4 Probe Station, Desert Cryogenics) and using a source meter (Keithley 2602), we measure the source-drain current ($I_{ds}$) as a function of the back gate voltage ($V_{gs}$) and source-drain bias ($V_{ds}$). Samples are typically kept under vacuum at 1x$10^{-6}$ Torr for 24 hours before measurement.

The focus of this report is the transport data presented in Fig.~\ref{ivs}, which shows a pronounced, notch-like drop in current over a small bias window in the vicinity of the broader and smoother conductance dip typical of an ambipolar FET. The notch was found to shift reproducibly with bias $V_{ds}$. Closer examination reveals trapping and detrapping events on both sides of the notch. The RTS signatures
are robust and detectable for several days under vacuum with clearly resolved amplitude at room temperature. \textcolor{black}{The} three-state RTS \textcolor{black}{is observed} in the electron conduction region (Fig.~\ref{ivs}), but \textcolor{black}{not for hole conduction} (except a `notch' with analogous gate-dependent shifts, that could be the remnants of a similar but narrow blockade window for hole conduction). While RTS is not unexpected in this type of device, its amplitude at ambient temperatures is very unusual. The truncated blockade with different RTS time constants suggests a multiple-trap effect that is electronically correlated. It is worth emphasizing that these signatures were not observed in an isolated sample, but on multiple CNT samples on separate chips. These observations motivated a comprehensive theoretical treatment of the data.

{\it{Theory.}} {The unique transfer characteristic ($I_{ds}-V_{gs}$)
is modeled} using the non-equilibrium Green's function (NEGF) formalism \cite{datta} coupled with a Monte Carlo simulation of the stochastic trap dynamics near resonance \cite{ieeesensors}. \textcolor{black}{As we establish later, the truncated blockade
window can be explained by invoking} two traps sitting inside the oxide between the CNT channel and backgate. Each trap density of states (DOS) is modeled by a Lorentzian of the form $D_t(E)={\gamma_t/2\pi}{[(E-\epsilon_t)^2+(\gamma_t/2)^2]}$, while the CNT DOS is given by $D_{ch}(E)=D_0 {|E|}\Theta(|E|-Eg/2)/{\sqrt{E^2-(E_g/2)^2+\eta}}$ \cite{guo}. $D_0= {8}/{3 \pi a_{cc}\gamma_0}$ is the low-bias metallic CNT DOS ($a_{cc}$ denotes the carbon bond length, $\gamma_0=2.5$ eV is the nearest-neighbor overlap integral, $E_g$ is the CNT band-gap, $\eta$ \textcolor{black}{is} a phenomenological dephasing parameter
\textcolor{black}{for incoherent scattering in the long tube}, and $\Theta$ is the Heaviside step function). The DOS states are shifted by the interatomic potentials to compute the charge and current in the NEGF formalism. We calculate the number of electrons of the $i^{th}$ level ({\it{ch}}, {\it{$T_1$}} and {\it{$T_2$}}, \textcolor{black}{respectively labeling the CNT} channel, trap 1 and trap 2) by integrating its DOS with the weighted Fermi functions $f_{S,D}$ ({\it{S}}: Source,
{\it{D}}: Drain). With contact broadening parameters $\gamma_{S,i}$ and $\gamma_{D,i}$ we obtain:
\begin{equation}
N_i=\int dE D_i(E-U_i)\biggl[\frac{\gamma_{S,i}f_S(E)+
\gamma_{D,i}f_D(E)}{\gamma_{S,i}+\gamma_{D,i}}\biggr], 
\label{Neqn}
\end{equation}

\textcolor{black}{While the NEGF model is widely used for computing current flow, the non-triviality in this case
lies in the physics of its ingredients, specifically, {{self-consistency}}, {{stochasticity}} and most significantly, {{self-interaction correction}}. The latter many-body problem, positing that a charge should not feel a potential due to itself, can be solved exactly for a few scattering ground states by introducing an explicit level-dependent potential. For example, the strongly correlated Coulomb potential acting on trap 1 is given by
$U_{T1}=\Delta N_{ch}U_{0,T1-ch}+\Delta N_{T2}U_{0,T1-T2}$,
where $\Delta N_i$ is the deviation in \textcolor{black}{the $i^{th}$ level} population from equilibrium, and $U_{0,j-i}$ is the Coulomb interaction matrix between the $i^{th}$ and $j^{th}$ levels. These potentials are calculated self-consistently with the computed charges from Eq.~\ref{Neqn}, as all 3 levels are affected by each other.
(We also need to invoke a sizeable Laplace asymmetry so that one trap can `passivate' another but not be passivated by
it. This is provided by the logarithmic potential variation perpendicular to the CNT, generating widely different capacitive gate transfer factors $\alpha_{g}$ for the distinct trap energies that slip at different rates past each other).} Finally, the converged channel potential yields the steady-state CNT current
\begin{equation}
I=\frac{2e}{h}\int dE D_{ch}(E-U_{ch})\frac{\gamma_{S,ch}\gamma_{D,ch}}{\gamma_{S,ch}+\gamma_{D,ch}}[f_S(E)-f_D(E)]
\end{equation}

In order to obtain the stochastic RTS signatures, the steady-state equations are modulated by the time-dependent trap potentials. These potentials become stochastic near the gate-tuned resonance between the trap's energy level and the CNT contact Fermi energies, so that the occupancy of the traps become fractional. To simulate the stochastic trap dynamics, we set up a two-state Monte Carlo simulation that depends only on the trap capture time $\tau_c$ and emission time $\tau_e$, relevant to a filled and an empty trap respectively \cite{ieeesensors}. For a trap with energy $\epsilon_T$ in thermal equilibrium with its environment, detailed balance requires \cite{ralls,uren,jiang1,jiang2}
\begin{equation}
\frac{\tau_c}{\tau_e}=2 \exp{\biggl[\frac{-(\epsilon_T-E_F)+\alpha_g |e|(V_{gs}-V_{g0})}{k_BT}\biggr]}
\label{balance} 
\end{equation}
The prefactor accounts for spin degeneracy, while $V_{gs}$ is the gate voltage relative to the grounded source. $V_{g0}$ is the value of the minimum in the ambipolar $I_{ds}$-$V_{gs}$ curve, which corresponds to a gate offset \cite{kim} due to other \textcolor{black}{distant} traps that do not contribute to the dynamics of the RTS signals. We use a given scan rate to convert temporal noise into voltage noise.

\begin{figure}[ht]
\hskip 0.2cm\centerline{\epsfxsize=2.8in\epsfbox{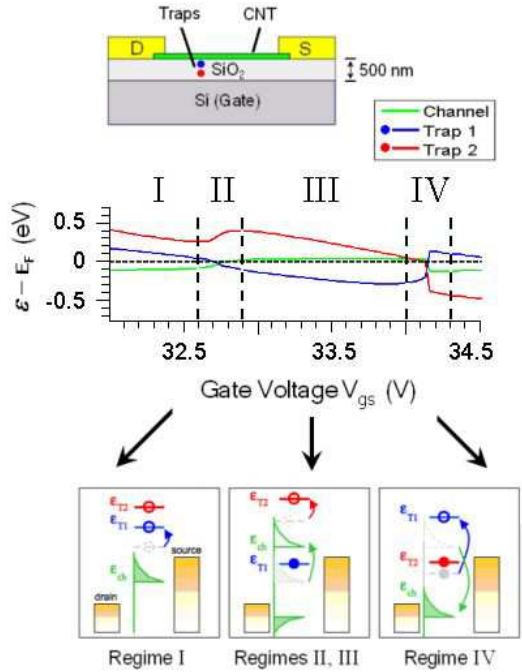}}
\vskip -0.4cm
\caption{{\it{(Top) Schematic diagram of CNT-FET with traps in $SiO_2$. (Center) Simulated evolution of trap levels and channel conduction band-edge under gate bias, influenced by self-consistent Coulomb repulsion. The sequence of energy level diagrams (below) show the exchange of level positions under these repulsive forces. Expulsion of the channel from the conduction window (regimes II, III) blocks the current, while its reintroduction (regime IV) unblocks it.}}}
\label{levels}
\end{figure}

{\it{Results.}} Fig.~\ref{ivs} demonstrates a compelling reproduction of the experimental data by our simulation model (parameters listed in \cite{param}). While the detailed transition voltages and current jumps depend on these parameters, the overall physics of blocking and unblocking is robust with respect to parameter variations. Our only assumption is
that the traps sit at different distances from the channel with a corresponding hierarchy of interaction strengths
($U_{0,T1-T2} > U_{0,T1-ch} > U_{0,T2-ch}$). The physics of the RTS-mediated blocking and unblocking can be understood by looking at the steady-state dynamics of the individual traps and channels computed by our model under their mutual self-consistent repulsion.

Fig.~\ref{levels} \textcolor{black}{plots} the gate-induced migration of the \textcolor{black}{levels (the peak DOS of the traps and the conduction band-edge for the CNT)}, \textcolor{black}{as extracted from their self-consistently computed steady-state potentials $U_i$}. The levels slip at very different rates controlled by the Laplace part $\alpha_g$ of the potential, varying logarithmically with distance from the gate. A positive gate voltage pushes the levels down, filling them sequentially upon entering the conduction window between the source and drain electrochemical potentials. Filling the CNT conduction band (regime I) repels trap 1 mildly and barely influences trap 2
sitting further away. When the first trap is filled, it pushes the CNT out of the conduction window and starts
the onset of blockade (regime II). This expulsion proceeds through an RTS sequence dictated by the steady-state occupancy of trap 1 (Eq.~\ref{balance}), ending with a fully filled trap and a completely blocked channel (regime III). At this stage, the differential slippage rates of the traps takes over, trap 2 overtaking trap 1 to get occupied (regime IV). Since trap 2 talks to trap 1 but not to the channel (recall the $U_0$ hierarchy), it expels the charge
on trap 1, thereby unblocking the channel (regimes IV, V). As before, regime IV proceeds through an RTS sequence (characterizing the second trap), thus creating the RTS-flanked blockade window seen experimentally and theoretically.

\begin{figure}[ht]
\hskip 0.2cm\centerline{\epsfxsize=2.5in\epsfbox{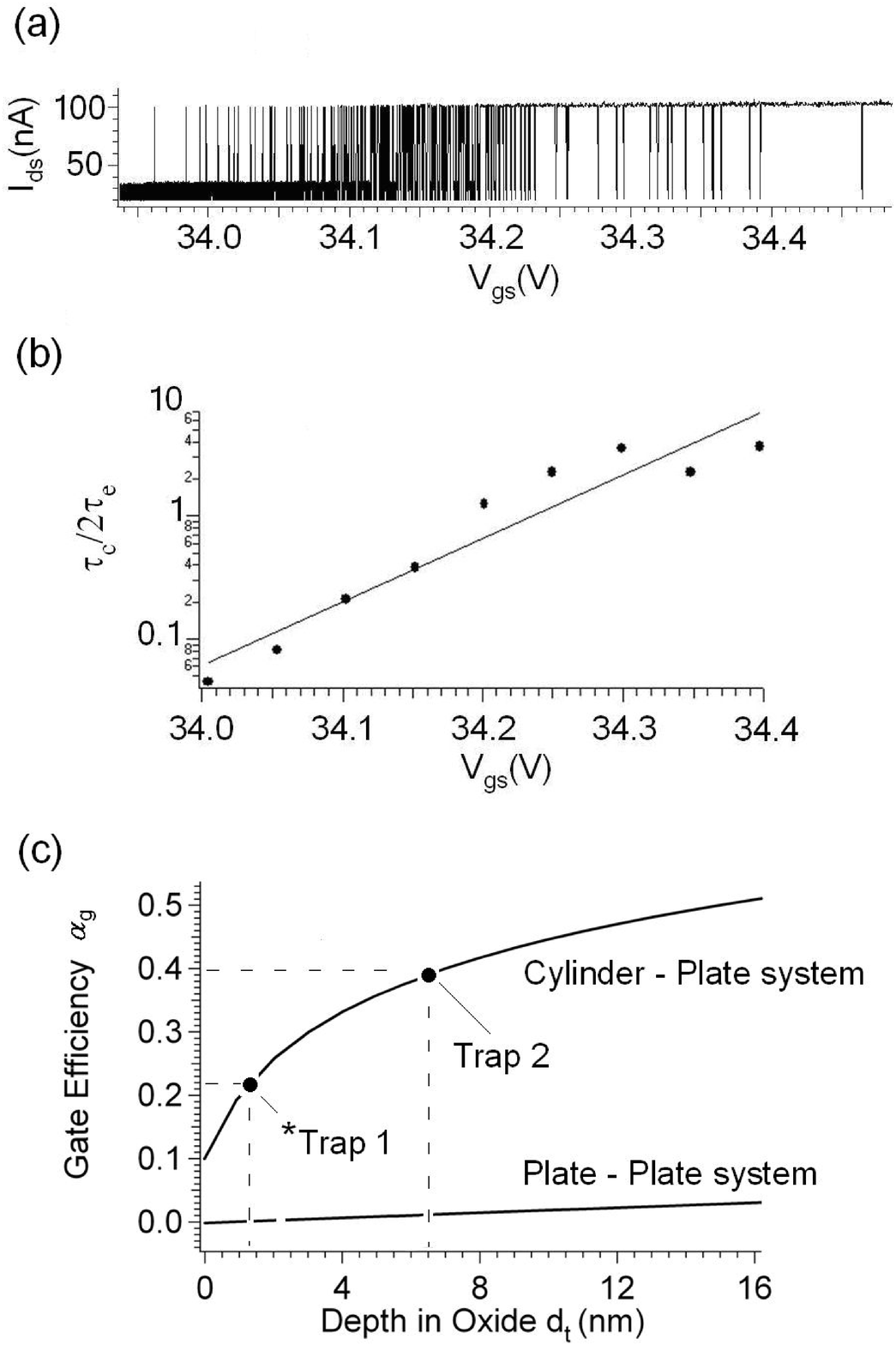}}
\vskip -0.4cm
\caption{{\it{(a) High resolution gate sweep showing RTS switching. Time interval between successive data points is 2 ms (Scan rate: 0.125 V/sec, resolution: 4000 data point/V). (b) Capture to emission time ratio vs $V_{gs}$ for regime IV. (c) Extraction of trap positions from experiment and simulation. }}}
\label{rts}
\end{figure}

The RTS signature was analyzed by taking high resolution scans in Regime IV \textcolor{black}{corresponding to the switching of trap 2} (Fig.~\ref{rts}a). Knowing the scan rate, \textcolor{black}{we can} extract information about $\tau_c$ and $\tau_e$ \textcolor{black}{for} trap 2, whose ratio is plotted as a function vs $V_{gs}$ in Fig.~\ref{rts}b. By Eq.~\ref{balance}, $\alpha_g$ is given by slope of the fitted line, which is found to be $0.40\pm 0.04$. \textcolor{black}{The} trap energy $\epsilon_{T2}$ can be obtained from the y-intercept, and is found to be $5.6 \pm 0.6$$ eV$ relative to the Fermi level. The high to low current switching region \textcolor{black}{characterizing the first trap} (Regime II) is not analyzed due to lack of \textcolor{black}{adequate} data points.
\textcolor{black}{The small operating drain bias} $(\sim 0.1$V), \textcolor{black}{relates} the gate capacitive factor $\alpha_g$ directly to an effective \textcolor{black}{trap} depth \textcolor{black}{$d_t$} under the oxide surface. The relationship is linear in the oxide for a regular MOSFET with planar geometry, $d_t\approx \alpha_gt_{ox}$, \textcolor{black}{$t_{ox}$ denoting} the thickness of the oxide layer \cite{celik1,celik2}. Because of the cylindrical structure of the carbon nanotube FET \textcolor{black}{the potential profile across} the oxide varies logarithmically with distance from the center of the conductor. The gate capacitance per unit length for a CNTFET is given by \textcolor{black}{$C = 2\pi\epsilon/\ln(2h/r)$}, where r is the nanotube radius, $h = t_{ox}+ r$ and \textcolor{black}{$\epsilon$} is the permittivity of the oxide. Using Eq.~\ref{balance} and the above capacitance, trap 2 is estimated to be 6.34 nm below the CNT in the oxide, consistent with typical values in the literature (\textcolor{black}{$\sim$} 0.2-20 nm) \cite{avouris3,zhou1,celik2}.
Fig.~\ref{rts}c shows the trap positions extracted from our simulation, assuming a logarithmically varying potential. For a linear potential, the significantly different $\alpha_g$s for unblocking require one of the traps to sit too far from the CNT to influence its current. The absolute necessity for fast varying potentials underscores the importance of low dimension to see these trap correlations.

The data reveal a rich spectrum of additional defect states, such as smaller current jumps at the bottom of the
blockade window, which we attribute to low-frequency device noise \cite{vandamme}. Furthermore, there is a gate offset $V_{g0}$ defining the minimum in the typical ambipolar \cite{kim} CNT-FET curve. The location of this minimum is independent of $V_{ds}$ and is shifted by 19.0 V. The observed blockade window shows an additional drift to lower gate voltage with time upon repeat scanning of $V_{gs}$ between 35 to 25 V over 18s intervals. The width of the window, however, remains the same between traces, so that each successive time trace can be superposed by a shift of 2V. Thermal treatment in $H_2$ for 4.5 hours at $200^{\circ}C$ followed by cooling and desiccation removes
the RTS but preserves the hysteretic shift, inconsistent with \cite{kim}. We tentatively attribute RTS to oxidative traps that can be passivated by $H_2$, while the shift in minimum $V_{g0}$ and the hysteretic effect may be
attributed to non-oxidative charge traps.

\textcolor{black}{In summary, we report unique transport dynamics whereby} one trap electrostatically `passivates' \textcolor{black}{another} without any chemical bonding between them, lifting a severe current blockade and fully restoring conduction. \textcolor{black}{The rapidly varying fields around the CNT enable us to perform detailed spectroscopy with fine resolution even at room temperature}, \textcolor{black}{allowing us to envision} a high precision `molecular barcode' for a wide range of surface adsorbates. Based on these observations, we suggest that it will be possible to engineer `receptor' states along a nanotube or similar 1-D channel that electrostatically couple to non-covalently-bound targets for detection with ultra-high specificity.

The authors are grateful to Chong Hu for assistance in using variable temperature probe station and Kamil Walczak for useful discussions. This work was partially supported by the NSF NIRT and CAREER awards.

\begin{thebibliography}{100}
\bibitem{param}The energy level positions used in the plots are  $\epsilon_{ch} = 1$ meV, $\epsilon_{T1}= 3$ eV and  $\epsilon_{T2}= 5.6$ eV relative to $E_F$. Broadenings $\gamma_{S,ch}=\gamma_{D,ch}=\gamma_{S,T1}= \gamma_{D,T1} = \gamma_{S,T2} = \gamma_{D,T2} = 50 $ meV. 
    Charging energies $U_{0,T1-ch} = 0.2$ eV, $U_{0,T1-T2} = 0.48$ eV and $U_{0,T2-ch} = 0$ eV.
\end{thebibliography}

\end{document}